\documentclass[aps,pra,twocolumn,superscriptaddress,
floatfix,reprint]{revtex4-1}
\usepackage{amsmath}
\usepackage[next]{inputenc}
\usepackage[dvips]{epsfig}
\usepackage{bbm,bm,bbold}
\usepackage{booktabs}
\usepackage{svg} 
\usepackage{multirow}
\usepackage{hhline}
\usepackage{comment}
\usepackage{float}
\usepackage{enumitem} 

\usepackage{amsmath,amsfonts,amssymb,amsthm}
\usepackage{color}
\usepackage{graphicx,latexsym}

\usepackage{microtype}

\usepackage[colorlinks=true,urlcolor=blue,citecolor=blue,linkcolor=blue]{hyperref}

\usepackage{lipsum}

\usepackage{nameref}
\usepackage{varioref}
\usepackage{cleveref}

\usepackage{natbib}

\usepackage{mathtools}

\DeclarePairedDelimiter\ket{\lvert}{\rangle}
\DeclarePairedDelimiterX\braket[2]{\langle}{\rangle}{#1\,\delimsize\vert\,\mathopen{}#2}

\begin{document} 
\renewcommand{\vec}{\mathbf}
\renewcommand{\Re}{\mathop{\mathrm{Re}}\nolimits}
\renewcommand{\Im}{\mathop{\mathrm{Im}}\nolimits}

\title{Chiral electronic network within skyrmionic lattice on topological insulator surfaces}

\author{Matteo Wilczak}
\affiliation{Department of Physics and Centre for Theory of Quantum Matter, University of Colorado, Boulder, Colorado 80309, USA
}

\author{Dmitry K. Efimkin}
\affiliation{School of Physics and Astronomy and ARC Centre of Excellence in Future Low-Energy Electronics Technologies, Monash University, Victoria 3800, Australia}

\author{Victor Gurarie}
\affiliation{Department of Physics and Centre for Theory of Quantum Matter, University of Colorado, Boulder, Colorado 80309, USA
}
\begin{abstract}
We consider a proximity effect between Dirac surface states of a topological insulator and the skyrmion phase of an insulating magnet. A single skyrmion results in the surface states having a chiral gapless mode confined to the perimeter of the skyrmion. For the lattice of skyrmions, the tunneling coupling between confined states leads to the formation of low energy bands delocalized across the whole system. We show that the structure of these bands can be investigated with the help of the phenomenological chiral network model with a kagome lattice geometry. While the network model by itself can be in a chiral Floquet phase unattainable without external periodic driving, we show how to use a procedure known as band reconstruction
to obtain the low energy bands of the electrons on the surface of the
topological insulator for which there is no external driving. 
We conclude that band reconstruction is essential for the broad class of network models recently introduced to describe the electronic properties of different nanostructures. 
\end{abstract}
\maketitle

\noindent
\section{Introduction}
Being intuitive and analytically tractable, tight-binding models form the foundation of electronic band theory and the topological classification of solids~\cite{TopClassification1,TopClassification2}. These models, however, do not extend naturally to certain twisted heterostructures, e.g. twisted biased bilayer graphene~\cite{NetworkTBGEfimkin, NetworkTBG1,NetworkTBG2,NetworkTBG3,NetworkTBG4,RecherTBG1,RecherTBG2,RecherTBG3}, which can be seen as moire mosaics composed of topologically distinct tiles. Their interfaces trap unidirectional and protected electronic states, ensuring that the low-energy physics of these heterostructures can be captured by network models, which phenomenologically describe the interface states. The topological classification of network models exhibits intriguing mathematical connections to periodically driven tight-binding models and recently attracted substantial attention~\cite{NetworkPhaseRotation,NetworkFloquetMapping,BandReconstruction1,PhotonicKagomeNetwork}
.

The existence of electronic networks has been predicted in twisted biased bilayer graphene~\cite{NetworkTBGEfimkin, NetworkTBGEfimkin, NetworkTBG1,NetworkTBG2,NetworkTBG3,NetworkTBG4,RecherTBG1,RecherTBG2,RecherTBG3}, double-aligned graphene-hexagonal boron nitride~\cite{KagomeNetworkFalkoTh}, periodically strained graphene~\cite{KagomeNetworkMele}, and commensurate collective ordered states in $1\hbox{T}-\hbox{TaS}_2$~\cite{NetworkCDW1}. While some of these predictions have already been confirmed via scanning tunneling~\cite{NetworkTBGExp1}, transport~\cite{KagomeNetworkFalkoExp,NetworkTBGExp3}, and plasmonic probes~\cite{NetworkTBGExp2}, the set of confirmed electronic networks is still sparse, and it would be highly desirable to extend this set.

\begin{figure}[b]
	\centering
    \begin{minipage}[htb]
   {0.28\textwidth}
    \vspace{0pt}
        \centering
\includegraphics[width=\textwidth]{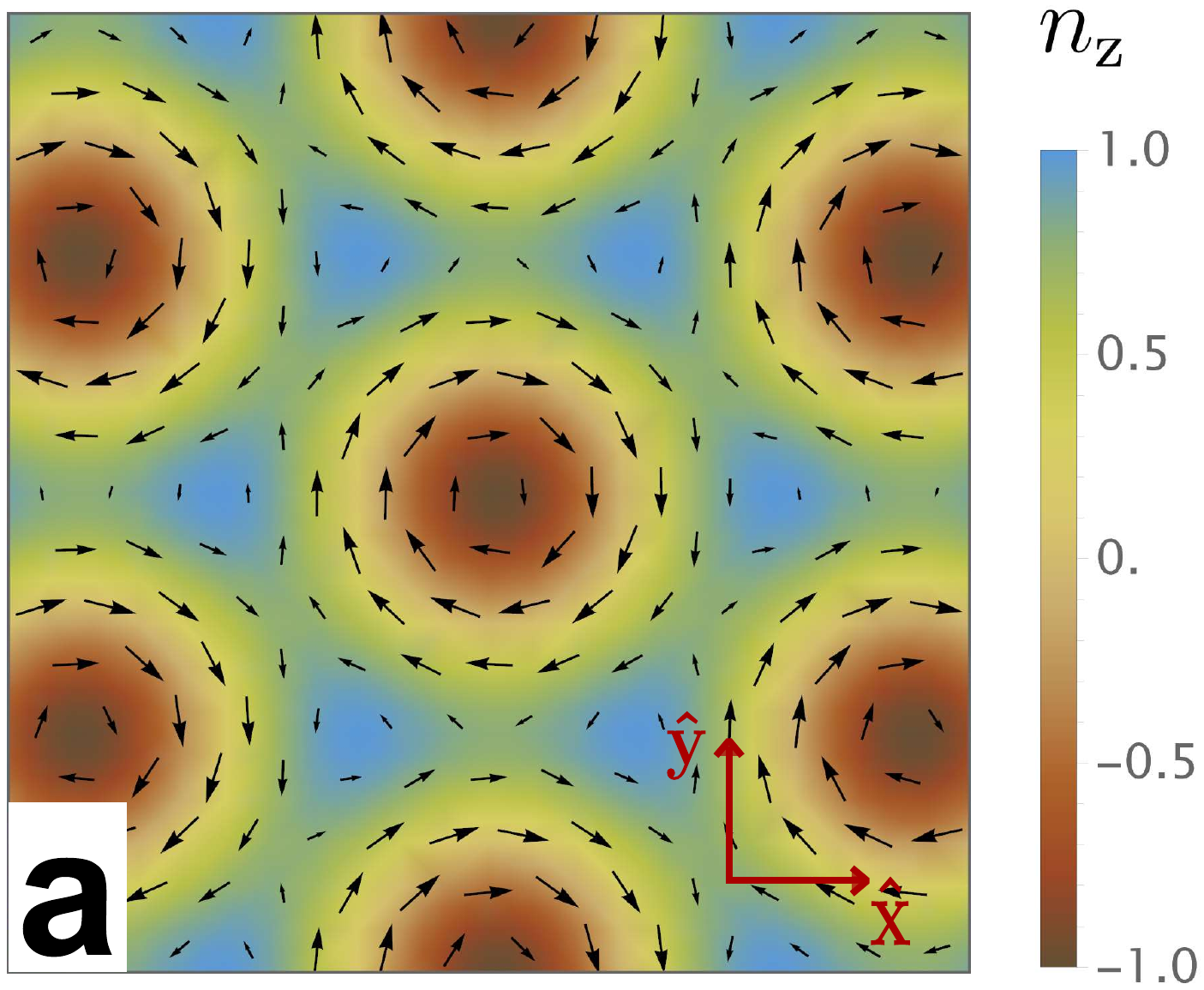}
    \end{minipage}
    \centering
    \begin{minipage}[htb]{0.17\textwidth}
    \vspace{0pt}
        \centering
\includegraphics[width=\textwidth]{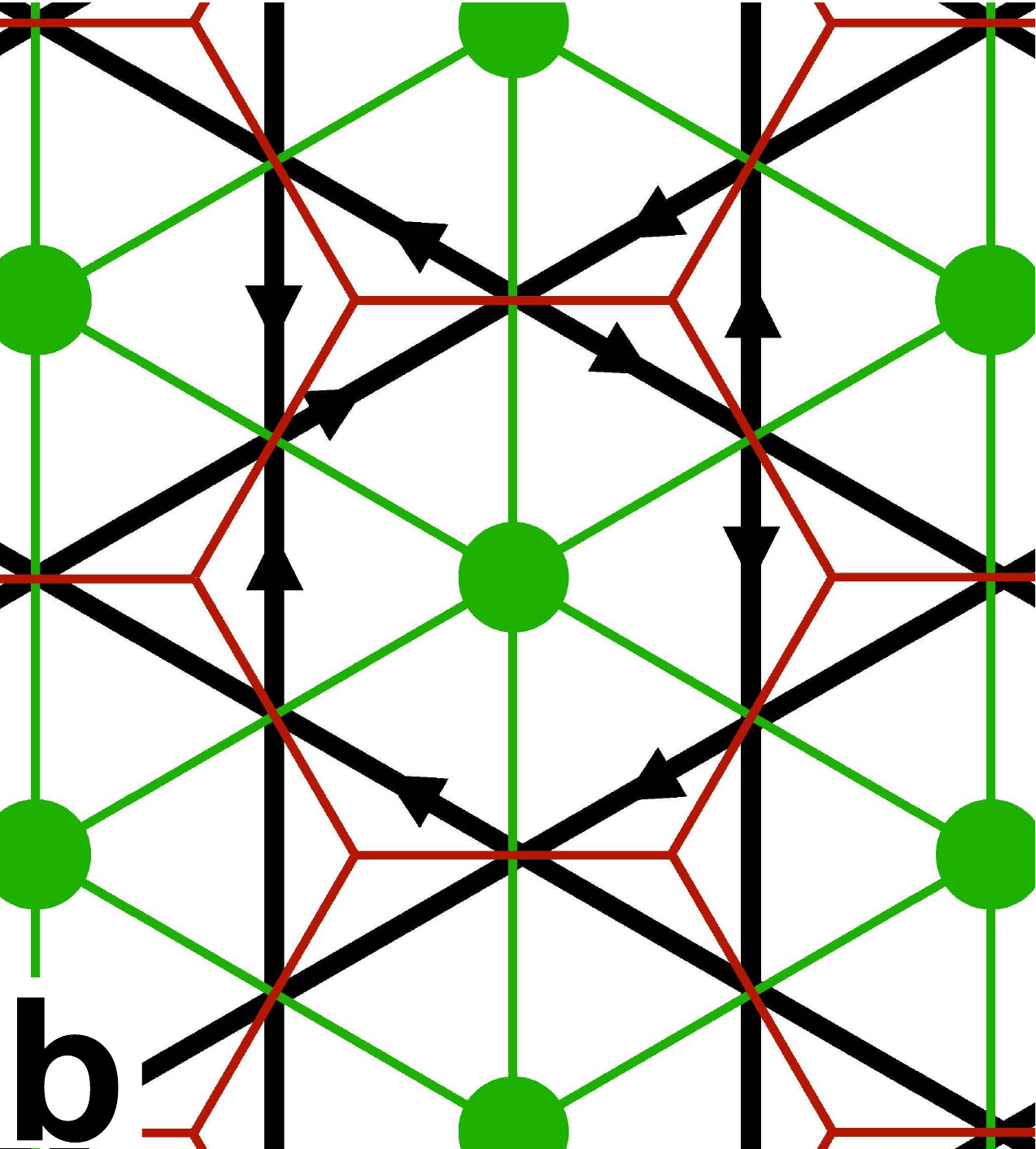}
   \end{minipage}
\caption{ (a) The spatial profile of the unit vector $\vec{n}(\vec{r})$ following the magnetization for the triangular skyrmionic lattice. (b) The Kagome network is formed by links describing chiral electrons (black arrows) circulating skyrmions (green dots) and scattering nodes (intersections of black lines) describing tunneling between chiral states trapped at adjacent skyrmions.}
\label{Fig1}
\end{figure}

Another promising platform for topological confinement engineering is the surface of a topological insulator hosting Dirac electronic states. Being robust to non-magnetic disorder, Dirac states are very sensitive to intrinsic or proximity-induced magnetic perturbations~\cite{MagneticProximity1,MagneticProximity2}. In particular, the exchange coupling with the out-of-surface component of magnetization generates the Dirac mass and opens a gap in the surface spectrum. The resulting state is topological, and its Chern number $\mathcal C=\pm 1/2$ depends on the sign of the Dirac mass. As a result, Dirac electrons experience topological confinement at the zero-mass lines induced by the magnetic domain walls~\cite{DW1,DW2,DW3,DW4,DW5} and skyrmions~\cite{TISkyrmionsEfimkin,TISkyrmions2,TISkyrmions3,TISkyrmions4,TISkyrmions5}. Furthermore, their exchange coupling with the triangular skyrmion lattice results in the formation of electronic minibands, which experience a series of topological transitions~\cite{TISkyrmionArun}. The corresponding magnetization profile, presented in Fig.~\ref{Fig1} (a), induces a mosaic composed of topologically distinct tiles, and a natural question arises of whether electronic minibands and their topologies can be described by a network model.   

In this paper, we revisit the coupling between Dirac electrons and the triangular skyrmion lattice at the surface of a topological insulator. We argue that the low-energy minibands and their topologies can be effectively captured by the Kagome network model, which is sketched in Fig.~\ref{Fig1} (b). The boundaries between the skyrmions form the honeycomb lattice
shown in red in Fig.~\ref{Fig1}. The Kagome lattice,
shown in black, arises as the medial lattice of that honeycomb lattice. Its links follow the zero-mass lines encircling individual skyrmions and are interrupted by the scattering nodes describing tunneling between states trapped at the adjacent skyrmions.  However, the na\"ive network model relying on the energy-independent scattering matrix only partially describes the minibands and their associated topological transitions. Additionally, one of the predicted network phases is the chiral Floquet phase, whose presence in time-independent systems we recently demonstrated to be self-contradictory~\cite{BandReconstruction1}. 
To resolve these issues, we utilize a calculation method referred to as band reconstruction, which accounts for the energy dependence of the network model parameters. These self-consistent calculations address the discrepancies, providing an accurate description of the electronic minibands and their topologies. Consequently, this confirms that the low-energy physics of the system can indeed be described by the chiral Kagome network model.

The rest of the paper is organized as follows. Sec.~\ref{sec: micromodel} is devoted to the microscopic theory of the proximity effect describing the Dirac electrons' interactions with individual skyrmions and with a triangular skyrmion lattice. In Sec.~\ref{sec:Network Model} we review the Kagome network model including its band structure and phase diagram and compare its predictions to the microscopic calculations. We further argue that the band reconstruction procedure is essential to properly capture the topological transitions of electronic minibands. Sec.~\ref{sec:Discussions} presents discussions and
conclusions.   

\noindent
\section{The microscopic model}
\label{sec: micromodel}
\subsection{The magnetic proximity effect}
Consider an interface between the surface of a topological insulator and a thin insulating magnetic film that hosts magnetic skyrmions. The exchange coupling between the Dirac surface states and the magnetic moments in the film can be parameterized by the energy $\Delta$ and described by the following Hamiltonian
\begin{equation}
H=v [\vec{p} \times \bm{\sigma}]_z - \Delta \; \bm{\sigma}  \cdot \vec{n}(\vec{r}). 
\end{equation}
Here, $v$ is the velocity of Dirac states, and the Hamilton acts on their spinor wave function, $\hat{\psi}(\vec{r})=\{\psi^\uparrow(\vec{r}),\psi^\downarrow(\vec{r})\}^T$. The unit vector $\vec{n}(\vec{r})$ follows the magnetization in the magnetic film, and the corresponding skyrmionic texture is sketched in Fig.~\ref{Fig1} (a). Interestingly, the Hamiltonian $H$ can be rewritten as follows
\begin{equation}
H=v \left[\left(\vec{p}- \frac{e}{c} \vec{a}(\vec{r})\right) \times \bm{\sigma}\right]_z - \Delta \; \sigma_z  \cdot \vec{n}_z (\vec{r}). 
\label{eq:Ham2}
\end{equation}
The out-of-plane component of the magnetization, $\vec{n}_z(\vec{r})$, induces a spatially dependent Dirac mass, while the in-plane component, $\vec{n}_{||}(\vec{r})$, generates the emergent electromagnetic field described by the vector potential $\vec{a}(\vec{r})=-c\Delta [{\hat {\vec e}}_\mathrm{z}\times \vec{n}_{||}(\vec{r})]/e v$.  
The static texture can produce an emergent magnetic field given by 
$B(\vec{r})= - c\Delta \, \mathrm{div} \, \vec{n}_{||}(\vec{r})/e v$  only if $\mathrm{div} \, \vec{n}_{||}(\vec{r})$ is nonzero. This unconventional coupling is unique to Dirac electrons and results in their rich interplay with topological magnetic defects~\cite{TISkyrmions3,TISkyrmions4, TISkyrmions5,TIRandomMFEfimkin} and spin wave excitations~\cite{TISpinWaves1,TISpinWaves2,TISpinWaves3}.

For a material realization, we consider the magnetic proximity effect between the surface of the TI $\hbox{Bi}_2 \hbox{Te}_3$ (or $\hbox{Bi}_2 \hbox{Se}_3$) and an insulating magnetic film of $\hbox{Cu}_2 \hbox{OSeO}_3$. The latter exhibits a triangular skyrmion lattice phase below the transition temperature $T_0 \approx 57 \; \hbox{K}$~\cite{SkyrmionFilm, SkyrmionReview1,SkyrmionReview2,SkyrmionReviewTretiakov}. The lattice period has been reported to be in the range $L \approx 63 \text{--} 73 \; \hbox{nm}$, which is several times larger than the typical skyrmion size, $r_\mathrm{s} \approx 25 \; \hbox{nm}$. Skyrmions are stabilized by Dzyaloshinskii-Moriya interactions~\footnote{The skyrmion lattice phase is stabilized by an interplay of Heisenberg and Dzyaloshinskii-Moriya interactions along with a weak external magnetic field.}, which favor the Bloch-type in-plane vortex textures where $\mathrm{div} \, \vec{n}_{||}(\vec{r}) = 0$. Consequently, the emergent magnetic field vanishes, and the skyrmion lattice affects Dirac electrons only through a spatially varying, periodic Dirac mass.

The electronic dynamics are governed by the topological confinement along zero-mass lines within individual skyrmions and by tunneling between adjacent skyrmions. For an exchange coupling strength of $\Delta \approx 20 \, \hbox{meV}$, achievable in magnetic heterostructures~\cite{MagneticProximity3,MagneticProximity4}, the characteristic electronic localization length $r_\mathrm{\Delta} = \hbar v / \Delta \approx 20 \, \hbox{nm}$ is slightly smaller than but comparable to the skyrmion radius, supporting this physical picture. To gain further insights into the resulting electronic structure of Dirac electrons within the skyrmion lattice phase, it is useful to first review how they interact with a single isolated skyrmion.
 
\subsection{A single isolated skyrmion}

For a single magnetic skyrmion, the out-of-plane component $n_\mathrm{z}(\vec{r})$ varies smoothly between $n_\mathrm{z}=\pm 1$ and vanishes at the skyrmion radius $r_\mathrm{s}$. The topological confinement supports a single, chiral electronic mode that circulates the skyrmion in a fixed direction.  If the centrifugal effects are negligibly small, semiclassical arguments can be applied, and the chiral mode is described by the Jackiw-Rebbi solution
 \begin{equation}
\label{ChiralState}
\psi_{p_{\phi}}(r,\phi)=\begin{pmatrix} e^{-i \phi/2} \\
-e^{i \phi/2}
\end{pmatrix}\exp\left[i \frac{p_{\phi} r_\mathrm{s}\phi}{\hbar} - \int_0^{r}  \frac{ n_\mathrm{z}(r') d r'}{  r_\mathrm{\Delta}}  \right],
\end{equation}
where $r$ and $\phi$ are polar coordinates centered at the position of a skyrmion,
 $p_\phi$ denotes the momentum along the zero-mass line within the skyrmion, and $r_\mathrm{\Delta} = \hbar v / \Delta$ represents the localization length of the chiral mode introduced above. Applying Bohr-Sommerfeld quantization to the cyclic motion along the ring yields the condition $ 2\pi r_\mathrm{s} p_{\phi} / \hbar + \phi_\mathrm{B} = 2\pi m$, where $m$ is the integer orbital quantum number. The Berry phase term, $\phi_{\mathrm{B}} = -\pi$, arises from the electron spin rotation when it propagates along the closed loop. This results in a spectrum given by $\epsilon_{j} = \epsilon_\mathrm{s} j$, where $j = m + 1/2$ is the total angular momentum quantum number for the Dirac state, and $\epsilon_\mathrm{s} = \hbar v / r_\mathrm{s}$ defines the energy separation between levels. 

The presence of trapped states can be verified through explicit calculation by solving Eq.~(\ref{eq:Ham2}) numerically. Due to the topological nature of confinement, the spectrum of these trapped states is relatively insensitive to specific details of the skyrmion profile. For simplicity, we adopt the smooth Gaussian ansatz 
$n^0_z(r) = 1 - 2 \, \mathrm{exp}[- \mathrm{ln}(2) \, r^2 / r_\mathrm{s}^2]$. The resulting spectrum, shown in Fig.~2, includes a pair of trapped states that are well separated from the continuum of unbound electronic states.

While the localization length $r_\Delta$ is comparable to the skyrmion radius $r_\mathrm{s}$, making centrifugal effects non-negligible, the spectrum agrees reasonably well with semiclassical predictions. At smaller skyrmion radii, $r_\mathrm{s} \lesssim r_\Delta / 2$, finite size effects cause the trapped states to move out of the gap as the centrifugal energy dominates. For skyrmions arranged in a triangular lattice, tunneling between trapped states in neighboring skyrmions leads to the formation of electronic minibands. 

\begin{figure}[t]
	\label{Fig5}
	\vspace{-2 pt}
	\includegraphics[width=7.9 cm]{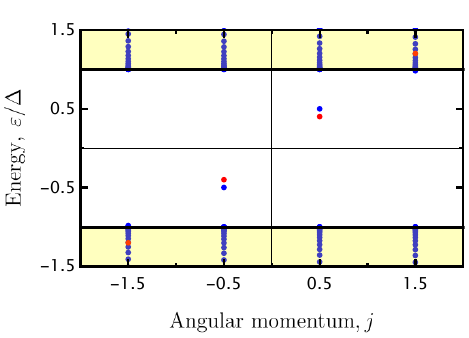}
	\vspace{-2 pt}
	\caption{ \label{fig:SingleskyrmionSpectrum}
    {The electronic spectrum of an isolated skyrmion, calculated numerically (blue) and analytically using semiclassical arguments (red). For the parameters considered, the skyrmion traps a pair of chiral states that are well-separated from the continuum of unbound states (yellow).}}
\end{figure}

\subsection{The triangular skyrmion lattice}
\begin{figure*}[htb]
	\centering
    \begin{minipage}[htb]
   {0.32\textwidth}
    \vspace{0pt}
        \centering
     \large{Trivial}
     \includegraphics[width=\textwidth]{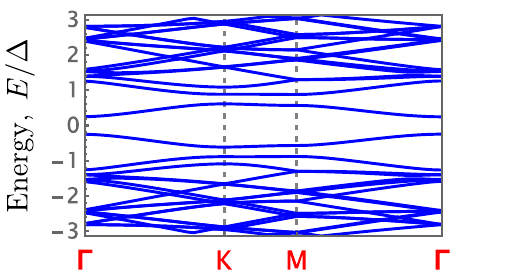} 
        \label{SkrymionLatticeBands_R=1.5_L=5.25_trivial_new}
    \end{minipage}\hfill
    \centering
    \begin{minipage}[htb]{0.32\textwidth}
    \vspace{0pt}
        \centering
    \large{Critical}
    \includegraphics[width=\textwidth]{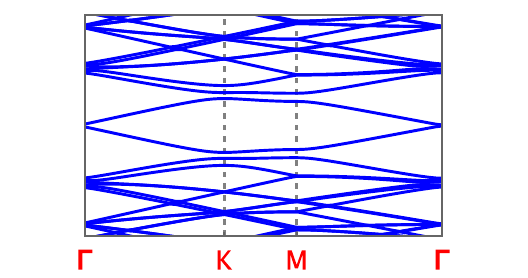} 
    \label{SkrymionLatticeBands_R=1.5_L=4.5_critical_new}
    \end{minipage}\hfill
    \centering
    \begin{minipage}[htb]{0.32\textwidth}
    \vspace{0pt}
        \centering
    \large{Topological}
    \includegraphics[width=\textwidth]{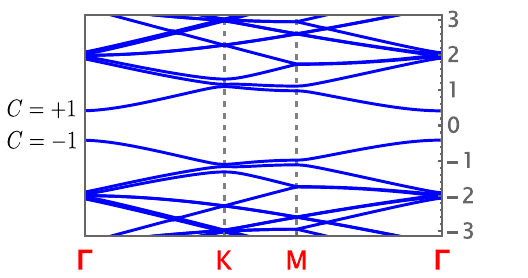} 
        \label{SkrymionLatticeBands_R=1.5_L=3.75_topological_new}
    \end{minipage}
    \hfill\\
    \centering
    \begin{minipage}[htb]{0.32\textwidth}
    \vspace{0pt}
        \centering
    \includegraphics[width=\textwidth]{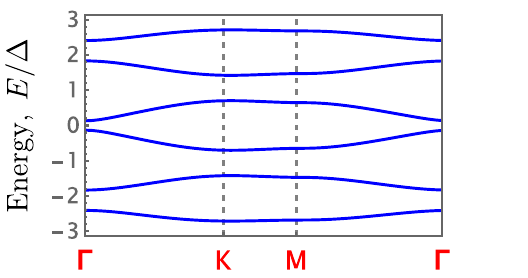} 
        \label{recbands_final_triv_r=0.25_b=0.15_new}
    \end{minipage}\hfill
    \centering
    \begin{minipage}[htb]{0.32\textwidth}
    \vspace{0pt}
        \centering
    \includegraphics[width=\textwidth]{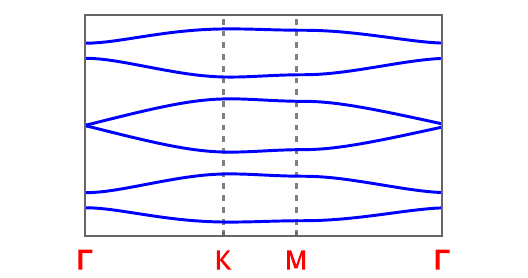} 
        \label{recbands_final_crit_r=0.33_b=0.15_new}
    \end{minipage}\hfill
    \centering
    \begin{minipage}[htb]{0.32\textwidth}
    \vspace{0pt}
        \centering
    \includegraphics[width=\textwidth]{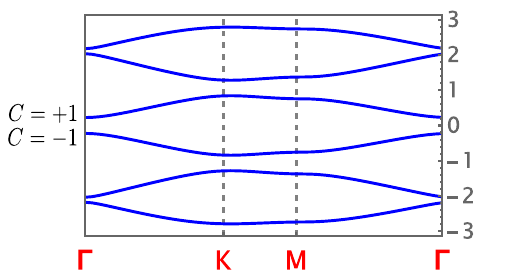} 
        \label{recbands_final_topo_r=0.48_b=0.15_new}
    \end{minipage}\hfill\\
    \textbf{Momentum}\caption{{Electronic minibands in the trivial phase (column 1), at the critical point (column 2), and in the Chern insulator phase (column 3) calculated using different approaches. Top row: microscopic calculations using the Dirac equation with skyrmion lattice period  $L = 95, 80, 65 \; \hbox{nm}$; Bottom row: phenomenological calculations using the network band reconstruction with an energy-dependent scattering angle parametrized by its maximal value $\theta_\mathrm{max} = \pi/8, \pi/6, \pi/4$.}}
 \label{fig:bands}
\end{figure*}

Individual skyrmions can exist as long-lived metastable states, but multiple skyrmions may also form an equilibrium phase, arranging themselves into a triangular lattice. The spatial distribution of the magnetization in this lattice configuration is illustrated in Fig.~1 (a), and the corresponding Dirac mass profile can be described by
\begin{equation}
n_z(r)=\sum_{i j} n^0_z(\vec{r}-\vec{r}_{i j }), \quad \quad \vec{r}_{i j} = i \vec{L}_1  + j \vec{L}_2  
\end{equation}
Here $\vec{L}_1=L ( \vec{e}_x  + \sqrt{3} \vec{e}_y)/2$ and $\vec{L}_2=L (- \vec{e}_x  + \sqrt {3}  \vec{e}_y)/2$ are translation vectors for the triangular lattice. In reciprocal space the corresponding eigenvalue problem defined by the Hamiltonian in Eq.~(\ref{eq:Ham2}), with ${\vec a}=0$, 
can be rewritten as   
\begin{equation}\label{eq:microscopiceval}
\begin{split}
v [(\vec{k}+\vec{G}_{ij}) \times \bm{\sigma}]_z \hat{\psi}_{ij}(\vec{k})+ \\ \frac{\Delta}{L^2} \sum_{i'j'} \sigma_z n^0_z(\vec{G}_{ij}-\vec{G}_{i'j'}) \hat{\psi}_{i'j'}(\vec{k})=E \hat{\psi}_{ij}(\vec{k}).
\end{split}
\end{equation}
Here, the $\bm{G}_{ij}$ are reciprocal lattice vectors, and $\vec{k}$ is confined to the first Brillouin zone. By truncating the reciprocal lattice vectors such that $|\vec{G}_{ij}| \leq G_{\text{max}}$, the eigenvalue problem can be solved through numerical diagonalization. The resulting spectra are shown in Fig.~\ref{fig:bands} (top row).

The resulting band structure depends on the skyrmion lattice spacing $L$. If the separation between zero-mass lines, $\Delta L = L - 2 r_\mathrm{s}$, for adjacent skyrmions is much larger than the localization length $r_\Delta$, tunneling between trapped states is minimal. In this case, the low-frequency bands are narrow and remain close to the energies of the trapped states, while the high-energy bands can be obtained through band folding within the nearly free electron model. As $\Delta L$ decreases, the lowest energy bands broaden and then begin to touch at $\varepsilon = 0$, undergoing a topological transition and acquiring nontrivial Chern numbers $\mathcal{C} = \pm 1$. The electronic minibands in the trivial, critical, and topological regimes are shown in Fig.~\ref{fig:bands} ($\Delta L/r_\mathrm{s}=2.25, 1.5, 0.75$ corresponds to $L = 95, 80, 65 \; \hbox{nm}$). The topological transition between them is the focus of the present paper. As $\Delta L$ decreases further, additional topological transitions involving higher energy bands can occur. However, the specific details of these transitions depend on the number of trapped states and their separation from the continuum of unbound electronic states, governed by the ratio $r_\mathrm{s} / r_\Delta$. Our results are consistent with theoretical predictions obtained using a different smooth profile to parameterize the skyrmion texture~\cite{TISkyrmionArun}.

\section{Kagome network Model}
\label{sec:Network Model}

\subsection{The network model}
The microscopic model introduced above provides comprehensive information about the electronic minibands, but it is excessive and demands extensive numerical computations. A simplified phenomenological model that focuses on capturing only the lowest energy minibands and their nontrivial topological properties would be highly desirable. Given the chiral nature of the trapped states, an appropriate model falls outside the traditional tight-binding framework and instead naturally resides within the network model class. In this model, chiral electron propagation along zero-energy lines is represented by directed links, while tunneling between adjacent trapped states is represented by scattering nodes. This class of models was initially developed in theories of the quantum Hall effect~\cite{Network1,Network2} and has recently been proposed to describe electronic nanostructures with engineered topological confinement~\cite{NetworkTBGEfimkin, NetworkTBGEfimkin, NetworkTBG1,NetworkTBG2,NetworkTBG3,NetworkTBG4,RecherTBG1,RecherTBG2,RecherTBG3,KagomeNetworkFalkoTh,KagomeNetworkMele,NetworkCDW1}.
\begin{figure}[t]
	\vspace{-2 pt}
	\includegraphics[width=7cm]{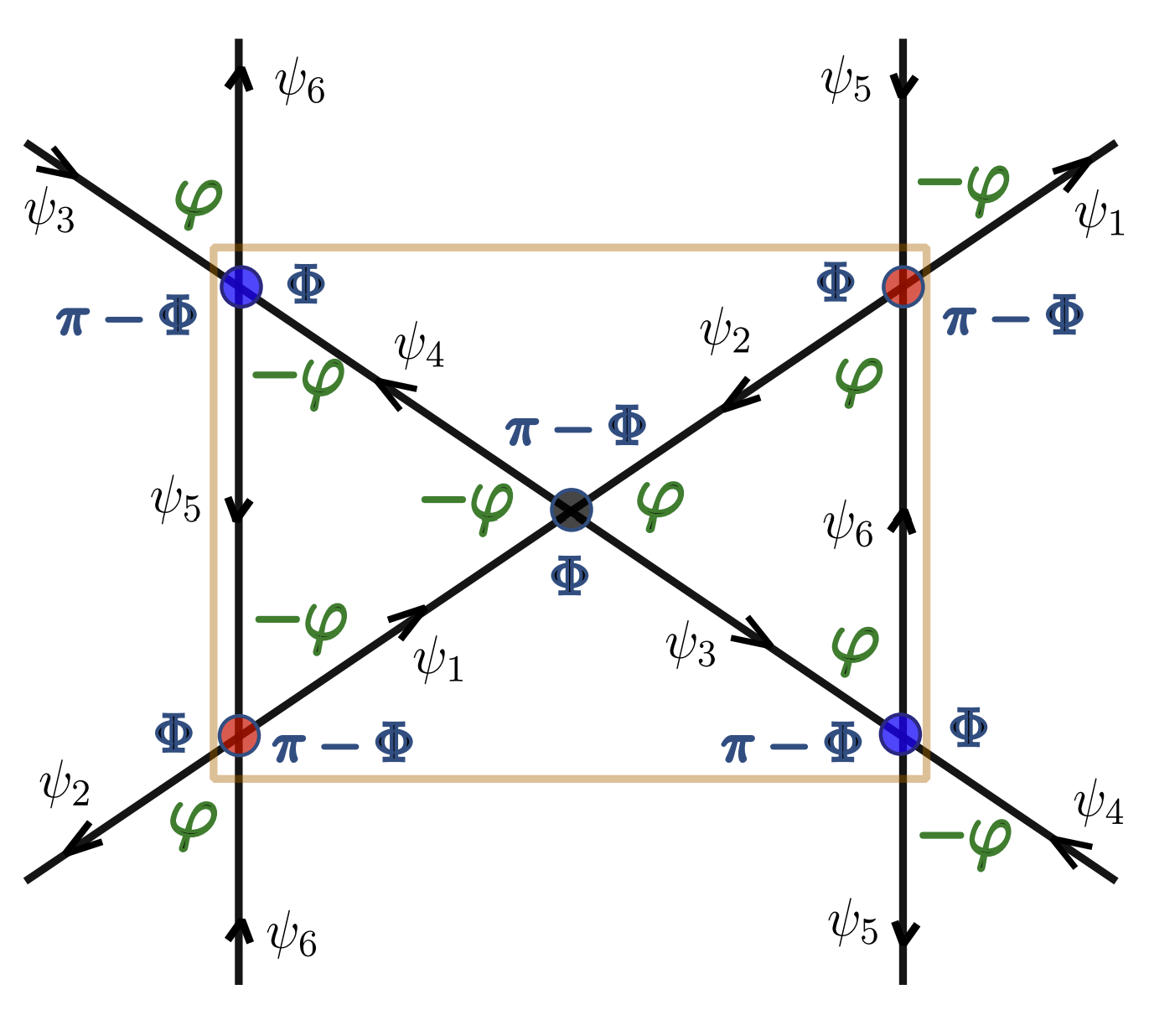}
	\vspace{-2 pt}
\caption{{The unit cell for the network model is denoted by the orange box. There is a single chiral state per link, and   $\psi_i$ ($i=1,... ,6$) is its amplitude. Extra phases picked up at the scattering nodes are shown in the corners.}\label{fig:unitcellwithphases}}
\end{figure}

For the triangular skyrmion lattice, the resulting network exhibits a Kagome geometry, as illustrated in Fig.~\ref{Fig1}~(b). Its unit cell takes the shape of a ``bow-tie," formed by two connected triangles shown in Fig.~\ref{fig:unitcellwithphases}. An electron propagates freely along six in-equivalent links, and its wave function has six components, i.e $\ket{\psi_{ij}} = (\psi^1_{ij}, \cdots, \psi^6_{ij})^T$~\footnote{The translation vectors can be chosen to be the same as for the triangular skyrmion pattern. They are given by $\vec{l}_1=L ( \vec{e}_x  + \sqrt{3} \vec{e}_y)/2$ and $\vec{l}_2=L (- \vec{e}_x  + \sqrt {3}  \vec{e}_y)/2$.}. The chiral nature of the electrons' motion is encoded in the linear dispersion relation $\epsilon_k=v k$. The unit cell also contains three nonequivalent nodes, marked by red, blue, and black dots in Fig.~\ref{fig:unitcellwithphases}. Due to the symmetry of the underlying triangular skyrmion lattice, each node can be characterized by an identical unitary scattering matrix, $\hat{S}$. This matrix connects two incoming and two outgoing electronic states at each node and can be parametrized by four independent parameters as 
as follows  
\begin{equation}
\label{MatrixT}
\hat{S} = e^{i\phi_\mathrm{T}}
\begin{pmatrix}
\cos \theta \;  e^{i \Phi} & \sin\theta\; e^{i \varphi} \\
\sin \theta \;  e^{-i \varphi} & -\cos{\theta} \; e^{-i \Phi} 
\end{pmatrix}
\end{equation}
Here, the angle $\theta$ ranges from $0$ to $\pi/2$ and characterizes the strength of tunneling between hexagons (skyrmions). We adopt the convention that $\theta = 0$ ($\theta = \pi/2$) corresponds to uninterrupted circulation around the network's hexagons surrounding the skyrmions (the triangles forming the ``bow-tie" shape). The three phases $\phi_\mathrm{T}$, $\varphi$, and $\Phi$ range between $0$ and $2 \pi$, and are intricately related to the fluxes through the network's plaquettes (see Appendix A for further details).

\begin{figure*}[htb]
\centering
    \begin{minipage}[htb]{0.32\textwidth}
    \vspace{0pt}
        \centering
    \large{Chiral Floquet}
    \includegraphics[width=\textwidth]{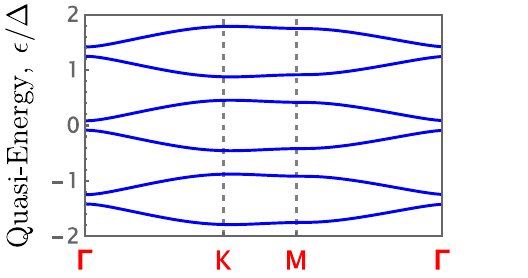} 
        \label{quasienergy_bandsfinal_0.25_triv_new}
    \end{minipage}\hfill
    \centering
    \begin{minipage}[htb]{0.32\textwidth}
    \vspace{0pt}
        \centering
    \large{Critical}
    \includegraphics[width=\textwidth]{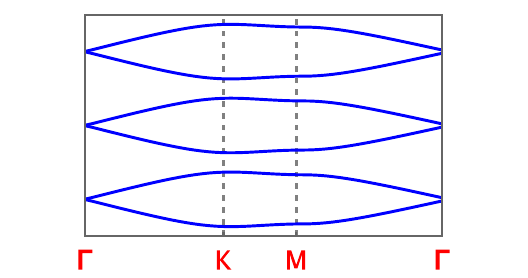} 
        \label{quasienergy_bandsfinal_0.33_crit_new}
    \end{minipage}\hfill
    \centering
    \begin{minipage}[htb]{0.32\textwidth}
    \vspace{0pt}
        \centering
    \large{Chern}
    \includegraphics[width=\textwidth]{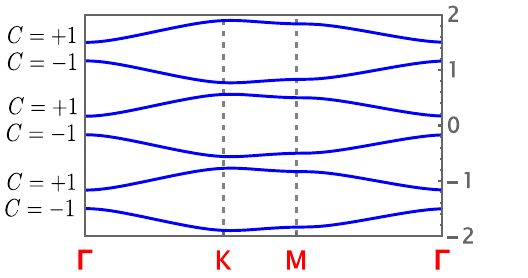} 
        \label{quasienergy_bandsfinal_0.5_topo_new}
    \end{minipage}\hfill\\
    \textbf{Momentum}\caption{{Phenomenological bands calculated using the Kagome network model with an energy-independent scattering angle $\theta = \pi/8, \pi/6, \pi/4$ corresponding to the chiral Floquet phase, critical point, and Chern insulator phase respectively. In the chiral Floquet phase, the topological winding number invariant is nonzero and all bands have zero Chern number.}}
 \label{fig:bandsnetwork}
\end{figure*}
The time evolution of the network model is 
captured by the 
``one-step" Floquet evolution operator $U_{\vec{k}}$ which consists of $2 \times 2$ blocks corresponding to scattering matrices and relates $\ket{\psi(t+1)}$ to $\ket{\psi(t)}$. 
 
As is standard in the Floquet dynamics, 
we need to solve the evolution operator
eigenvalue problem:
\begin{equation}
\hat{U}_\vec{k}  \ket{\psi_\vec{k}} =  e^{-i \varepsilon} \ket{\psi_\vec{k}}, \quad \quad \varepsilon=\frac{\pi r_\mathrm{s}}{3 \hbar v} E.  
\label{EigenvalueProblemU}
\end{equation}
Here $2\pi r_s/(6 v)$ is the time duration of the single Floquet step. 
The $6 \times 6$ matrix $\hat{U}_\vec{k}$ is given by
\begin{align}
\label{eq:TimeEvolutionU}
    \hat{U}_\vec{k} = \begin{pmatrix}
        0 & 0 & \sigma^x P_{\vec{k}2} \hat{S} P_{\vec{k}2} \\
        \hat{S} & 0 & 0\\
        0 & \sigma^x \hat{P}_{\vec{k}1}\hat{S} \hat{P}_{\vec{k}1}  \sigma^x & 0
    \end{pmatrix}
\vspace{1em}
\end{align}
where $P_{\vec{k} 1(2)}=\mathrm{diag}[e^{i \vec{k} \cdot {\vec L}_{1(2)}/2}, e^{-i \vec{k} \cdot {\vec L}_{1(2)}/2}]$. The resulting electronic spectrum is periodic, so the eigenvalues are referred to as quasienergies, and they are given by
\begin{equation}
E_{n m \vec{k}} = \frac{3 \hbar v}{\pi r_\mathrm{s} } ( \varepsilon_{n \vec{k}}+2\pi m),
\label{Qausienergies}
\end{equation}
where $\varepsilon_{n \vec{k}}$ is $n$-th eigenvalue ($n=1,..., 6$) of the operator $i \ln [\hat{U}_\vec{k}]$ and $m$ is an integer. Due to the block-diagonal nature of the matrix $\hat{U}_\vec{k}$ there is a phase rotation symmetry \cite{NetworkPhaseRotation}, which reduces the fundamental domain to $2\pi/3$ and the spectrum can be obtained by the periodic extension of just two bands. The periodic nature of quasienergies hints at a mathematical connection to the stroboscopic evolution of periodically driven quantum systems~\cite{RudnerChiralFloquet}.

\begin{figure}[b]
\includegraphics[width=7.9 cm]{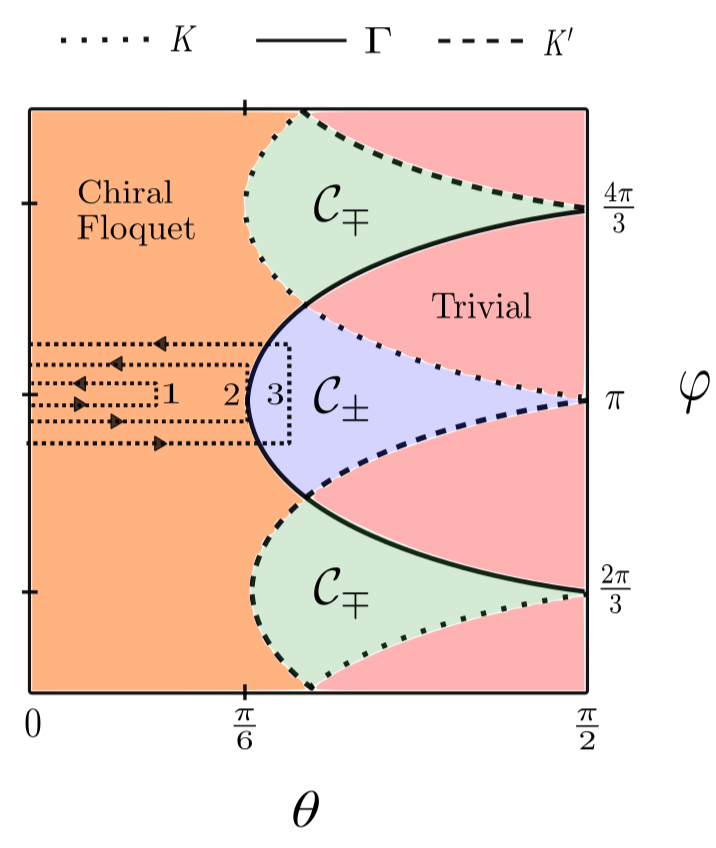}
	\vspace{-2 pt}
\caption{{The phase diagram for the network model with trivial, Chern, and chiral Floquet phases. The points of the Brillouin zone at which quasienergy bandgaps close are represented by dashed, dotted, or solid lines at each phase boundary. The network band reconstruction requires a choice of a path through the phase diagram. Three paths, specified by $\varphi = \pi$ and $\theta(E)$,  are shown by small dotted lines and are drawn displaced from the line $\phi=\pi$ for clarity.}}\label{fig:phasediagram}
\end{figure}

The evolution operator $\hat{U}_{\vec{k}}$ depends on four independent parameters: an angle $\theta$ and three phases $\phi_\mathrm{T}$, $\varphi$, and $\Phi$, however, not all of them are equally important. The scattering phases $\phi_\mathrm{T}$ and $\Phi$ can be eliminated by a uniform shift in both energy and momentum and are therefore irrelevant. Consequently, the phase diagram of the network model, which was previously discussed in \cite{PhotonicKagomeNetwork}, is governed by the angle $\theta$ and the phase $\varphi$ and is shown in Fig.~\ref{fig:phasediagram}, featuring three phases: trivial, Chern insulator, and chiral Floquet. The Chern phase is characterized by bands with nonzero Chern numbers and edge states that circulate around the network boundary. The chiral Floquet phase, characterized by a topological winding number~\cite{NetworkRLBL}, can also have chiral edge states despite all bands having Chern numbers equal to zero ~\footnote{It is important to note that the distinction between the chiral Floquet and trivial phases depends on the boundary choice, as pointed out in Refs.\cite{NetworkPhaseRotation, BandReconstruction1}. This can be verified graphically by looking at how modes are allowed to propagate along the edge of the lattice or by the fact that the value of the winding number which characterizes the chiral Floquet phase may be different for different choices of unit cell. Also, the winding number which characterizes the chiral Floquet phase cannot be calculated directly from $U_{\vec{k}}$ so this operator does not contain complete information about the topologies and one must instead construct some periodic time-dependent $U(t)$ with period $T$ such that $U(T) = U_{\vec{k}}$.}. All topological transitions between phases are accompanied by a series of band inversions occurring at the $\Gamma$, $K$, and $K'$ points of the Brillouin zone, as also illustrated in Fig.~\ref{fig:phasediagram}.

The network's electronic structure and topological properties are highly intricate. A natural question is how to select the scattering parameters $\theta$ and $\varphi$ so that the network model effectively captures the low-energy physics of Dirac electrons interacting with the triangular skyrmion lattice. First, we notice that the microscopically calculated electronic energies at the $K$ and $K'$ points match, indicating reflection symmetry. Additionally, the band inversion occurs at the $\Gamma$ point. From the network model's perspective, quasienergies at the $K$ and $K'$ points match only if the two types of triangles in the bow-tie are pierced by equal fluxes, i.e., $3\varphi = -3\varphi$. This condition restricts $\varphi$ to integer multiples of $2\pi/6$. For the network model's band gaps to close at the $\Gamma$ point, as in the microscopic model, $\varphi$ must be further restricted to values of either $0$ or $\pi$. Finally, to ensure that the gap between a pair of bands closes at a quasi-energy $E = 0$, the only feasible value for $\varphi$ is $\pi$.

The corresponding section of the phase diagram includes two phases: the Chern phase and the chiral Floquet phase. The topological transition between these phases occurs at $\theta_{c} = \pi/6$ and is accompanied by band inversion at the $\Gamma$ point.  The calculated electronic bands, shown in Fig.~\ref{fig:bandsnetwork}, have a similar shape and bandwidth to the microscopically calculated bands in Fig.~\ref{fig:bands}, however, at the critical point only the two lowest-energy microscopic bands experience a topological transition whereas every pair of network bands does. Furthermore, the network model includes a chiral Floquet phase which does not exist in the skyrmion lattice because it is not a periodically driven system. 

These significant differences between the network bands and the microscopically calculated bands indicate that the network model on its own as described in Eqs.(\ref{MatrixT})- (\ref{Qausienergies}) is insufficient to capture the physics of the low-energy minibands of the skyrmion lattice. To resolve this disagreement, we must move beyond the implicit assumption that the scattering matrix $\hat{S}$ is independent of the energy of electrons propagating through the network.

\subsection{Band Reconstruction}
In discussing the network model above, we have assumed that the scattering parameters are energy-independent. This led to network bands which do not match the microscopically calculated skyrmion lattice bands and their topological transition. By allowing the scattering parameters $\theta, \varphi$ of the newtork  to depend on energy we obtain bands that do. 

While the phase $\varphi = \pi$ is fixed by reflection symmetry, the scattering amplitude $\theta(E)$ is expected to exhibit a strong energy dependence. Tunneling becomes effective only when the incoming electron energy is comparable to the hybridization energy, which is determined by the overlap of wave functions for states localized at adjacent skyrmions. If the scattering amplitude $\theta(E)$ depends on energy, then Eq.~(\ref{Qausienergies}) extends to
\begin{align}
E_{n m \vec{k}} = \frac{3 \hbar v}{\pi r_\mathrm{s} }(\varepsilon_{n\vec{k}}\left(\theta(E_{n m \vec{k}}) \right) + 2 \pi m)
\end{align}
and is a nonlinear equation for the electronic dispersion $E_{n m \vec{k}}$. 
This procedure for self-consistently calculating the energy bands has recently been dubbed the network band reconstruction~\cite{BandReconstruction1,BandReconstruction2}. 

The energy dependence of the scattering amplitude $\theta(E)$ may be approximated by a model of two parallel zero-mass lines coupled by a tunneling barrier (see Appendix \ref{app:Eofscattering} for more details). Assuming the barrier is smooth, and its width is approximately half the length of each link in the network,  the energy dependence can be reasonably approximated as
\begin{equation}
\label{eq:thetaE}
\theta(E) =\frac{\theta_\mathrm{max}}{1+\left(
\pi r_\mathrm{s} E/8 \hbar v\right)^2}. 
\end{equation}
Here, $\theta_{\mathrm{max}}$ represents the maximum scattering angle, which is realized by electrons incident on the tunneling barrier with zero energy and is determined by the separation between zero-mass lines for adjacent skyrmions $\Delta L$.

For the energy-dependent scattering angle $\theta(E)$, the topology of electronic minibands is characterized by a path through the phase diagram instead of a point. Three paths are presented in  Fig.~\ref{fig:phasediagram}, indicated by dotted lines drawn displaced from the line $\varphi = \pi$ for clarity. Each path connects $\theta(-\infty)=0$  with $\theta(0)=\theta_\mathrm{max}$ and then returns back, i.e. $\theta(\infty)=0$. Nontrivial topologies of minibands are only possible if the path enters the Chern phase, which implies $\theta_\mathrm{max}>\theta_\mathrm{c}$.

The electronic bands calculated using the network band reconstruction are presented in Fig.~\ref{fig:bands} (bottom row). The low-energy behavior of the scattering angle can be approximated to be energy independent, $\theta(E)\approx \theta(0)$, which is why the network model with energy-independent scattering angle describes both shapes of the lowest energy pair of electronic minibands as well as their topologies. This is not the case for the other minibands.         

The topological numbers for minibands can be evaluated explicitly using the electronic wave functions calculated within the band reconstruction procedure. However, if the energy dependence of the scattering angle is smooth enough (this is the case for the set of parameters considered ), they can be understood in a simpler and physically transparent manner. Since the scattering angle depends on the energy, each of the reconstructed bands $E_{nm \vec{k}}$  corresponds to a particular range of scattering angles. If the bandwidths are small relative to $(\partial \theta / \partial E)^{-1}$ then we may approximate $\theta_{nm} \approx \theta(E_{nm \vec{k}})$. For $\theta_{max} < \theta_c$, all $\theta_{nm}$ take values in the chiral Floquet phase, and all reconstructed bands have vanishing Chern numbers. For $\theta_{max} > \theta_c$, some of the $\theta_{nm}$ may take values in the Chern insulator phase and the corresponding reconstructed bands $E_{nm \vec{k}}$ are then characterized by nonzero Chern numbers. Since $\theta(E)$ is peaked at $E=0$ and decays monotonically as $E \to \pm \infty$, initially only the pair of bands closest to $E=0$ undergoes a topological transition. As $\theta_{max}$ is tuned towards its maximal value $\pi/2$, higher energy bands also undergo topological transitions.
 This corresponds to a decreased separation between skyrmions $\Delta L$ or an increased coupling between Dirac electrons. If an individual skyrmion traps a fixed number of chiral states, the topological transition of the second pair of bands (top two bands and bottom two bands) does not happen simultaneously with the transition involving the lowest pair of bands, but at larger $\theta_\mathrm{max}$. This behavior is consistent with the microscopic calculations.

\section{Discussion}
\label{sec:Discussions}
The low-energy electronic states in double-aligned graphene-hexagonal boron nitride~\cite{KagomeNetworkFalkoTh} and periodically strained graphene~\cite{KagomeNetworkMele} have been argued to be described by the Kagome network model considered in this work but with different scattering parameters. Each topological interface of these heterostructures traps a pair of helical counter-propagating modes formed by electronic states in the vicinity of two inequivalent corners (K and K$'$) of the first Brillouin zone. As a result, the symmetry-based restrictions, $\phi=4\pi/3$ ($\phi=2\pi/3$) for the valley K (K$'$), are different from those for the considered chiral electronic network, $\phi =\pi$. 

For a material realization, we considered the magnetic proximity effect between the surface of the TI $\hbox{Bi}_2 \hbox{Te}_3$ (or $\hbox{Bi}_2 \hbox{Se}_3$) and an insulating magnetic film of $\hbox{Cu}_2 \hbox{OSeO}_3$. Coupling between Dirac electrons and a skyrmionic lattice can also be induced at the TI surface through a misaligned antiferromagnetic substrate such as $\hbox{CrI}_3$~\cite{TISkyrmionLatticeSubstrate} or by twisting a few-layer-thick magnetic TI, such as $\hbox{Mn}\hbox{Bi}_2 \hbox{Te}_4$~\cite{TISkyrmionLatticeTwisted1}. Alternatively, a skyrmionic spin-density wave state can be intrinsic to Dirac electrons, stabilized by the interplay between Coulomb repulsion and hexagonal warping~\cite{TISkyrmionLatticeInternal1, TISkyrmionLatticeInternal2}.  

The triangular skyrmion lattice phase in $\mathrm{Cu}_2 \mathrm{OSeO}_3$ is stabilized by the interplay of Heisenberg and Dzyaloshinskii-Moriya interactions under a weak magnetic field, $B \approx 0.05 \sim 0.2 \; \mathrm{T}$~\cite{SkyrmionFilm}, which has been disregarded thus far. The corresponding magnetic length lies in the range $l_\mathrm{B} \approx 58 \sim 112 \; \mathrm{nm}$ and exceeds both the electronic localization length, $r_\Delta \approx 20 \; \mathrm{nm}$, and the skyrmion size, $r_\mathrm{s} \approx 25 \; \mathrm{nm}$. Consequently, the magnetic field has a negligible effect on the trapping of electronic states at individual skyrmions. Its influence on miniband formation is characterized by the ratio $\Phi / \Phi_0 \approx 0.05 \sim 0.2$, where $\Phi$ is the magnetic flux through the triangular lattice unit cell of the skyrmionic lattice and $\Phi_0= \hbar c/e$ is the flux quantum. The neglection of the magnetic field is justified only for values close to the lower critical field, $B \gtrsim 0.05 \; \mathrm{T}$. For $B \lesssim 0.2 \; \mathrm{T}$, however, the magnetic field could induce magnetic oscillations akin to the precursors of the Hofstadter butterfly effect~\cite{NetworkTBGDasSarma}, which are beyond the scope of this paper. Additionally, the presence of a magnetic field is not required for other material realizations discussed above.   

Above the transition temperature $T_0 \approx 57 \; \mathrm{K}$~\cite{SkyrmionFilm}, the skyrmion lattice in $\mathrm{Cu}_2 \mathrm{OSeO}_3$ does not vanish but instead melts into a skyrmion liquid phase~\cite{SkyrmionMelting1,SkyrmionMelting2,SkyrmionMelting3}. While the size of individual skyrmions and their local triangular arrangement remain relatively intact, the distances between adjacent skyrmions become increasingly irregular as the temperature increases. The Kagome network model can still describe electron trapping and hopping between skyrmions in this phase, though with random scattering parameters, similar to the original Chalker-Coddington models~\cite{Network1,Network2}, which were introduced to describe localization-delocalization transitions in quantum Hall effect setups.

The formation of minibands in the Dirac surface state spectrum can be probed directly through ARPES and optical absorption, or indirectly via transport measurements. The quantization of Hall conductivity due to the nontrivial topology of electronic bands results in the anomalous Hall response, which can also be revealed in transport probes and in Faraday and Kerr magneto-optical effects~\cite{TISkyrmionLatticeOptics1,TISkyrmionLatticeOptics2}. Typically, nontrivial topology manifests itself via chiral and protected edge states. However, the presence of these states also requires a global energy gap on both sides of the edge. If the skyrmion lattice terminates at the magnetic film edge, the gapless nature of Dirac electrons outside the film does not favor any edge states. If a skyrmionic pattern is surrounded by a region with uniform out-of-plane ferromagnetic spin ordering, the interface supports a single chiral state connecting the topological minibands.  

To conclude, we have confirmed that proximity-induced coupling between the Dirac states of a topological insulator and magnetic triangular skyrmion lattice leads to the formation of electronic minibands with nontrivial topologies. Furthermore, these minibands and their topological transitions can be reproduced with the help of the phenomenological chiral network model with a kagome lattice geometry. Its proper topological classification requires the use of the band reconstruction procedure, which is also essential for other network models introduced to describe electronic nanostructures with engineered topological confinement. 
\section*{Acknowledgements}

We acknowledge support from the Australian Research Council Centre of Excellence in Future Low-Energy Electronics Technologies (CE170100039).  This work was also supported
by the Simons Collaboration on Ultra-Quantum Matter,
which is a grant from the Simons Foundation (651440,
VG, MW). 
We would like to thank Itamar Kimchi and Nigel Cooper for helpful discussions.

\bibliography{References}

\appendix

\section{Parametrization of the scattering matrix}
In this appendix we provide additional details about the phases $\phi_T, \varphi$ and $\Phi$ and their relationship to the fluxes through the plaquettues of the kagome network. When an electron is scattered around a particular corner it picks up the phase associated with that corner as in Fig. \ref{fig:unitcellwithphases}. The choice of which phases get assigned to which corners is a matter of convention and is specified by Eqs. \ref{MatrixT}, \ref{eq:TimeEvolutionU}.
With this choice of convention, electrons pick up a total phase of $\pi$ after circulating a hexagon in the network. This notably does not depend on the phase $\Phi$ since for each corner of the hexagon with a phase $\Phi$ there is another corner with phase $\pi-\Phi$.
When an electron goes around a triangle, however, it accumulates a phase of $\pm 3 \varphi$ where the sign depends on which of the two triangles of the unit cell the electron circulates. As a result, the phase diagram of the Kagome network model depends not only on $\theta$ but also on $\varphi$, while being independent of $\Phi$ and $\phi_\mathrm{T}$.

If we require the two types of triangles to have the same flux this imposes a condition on the parameter $\varphi = m\pi/3$ for some integer $m$. The paths through the phase diagram in Fig. \ref{fig:phasediagram} correspond to $m=3$.

\section{Energy Dependence of Scattering between two chiral wires}
\label{app:Eofscattering}
In order to better understand how the scattering between skyrmions depends on energy, we calculate the transmission and reflection amplitudes for a pair of counter-propagating chiral wires coupled by a box potential which has spatial range $2d$. To proceed we solve the following scattering problem
\begin{equation} - i v \frac{\partial \psi_R}{\partial x} + t(x) \psi_L = E \psi_R, \end{equation}
\begin{equation} i v {\frac{\partial \psi_L}{\partial x}} + t(x) \psi_R = E \psi_L.
\end{equation}
where $t(x)$ is the box potential of width $2d$ and strength $t$
\begin{equation} t(x) = t \, \theta(d-x) \theta(x+d).
\end{equation}
Focus first on $E>t$ or $E<-t$. 
Consider an incoming particle from the left. We find:
\begin{align} x<-d:& \  \left(  \begin{matrix}  \psi_R \cr \psi_L  \end{matrix}  \right) = \left( \begin{matrix}   e^{i E x/v} \cr R e^{-i E x/v}  \end{matrix} \right),\\[0.5em]
 x>d:& \  \left(  \begin{matrix}  \psi_R \cr \psi_L  \end{matrix}  \right) = \left( \begin{matrix}  T  e^{i E x/v} \cr 0  \end{matrix} \right),\\[0.7em]
 \begin{split}
 -d<x<d:& \  \left(  \begin{matrix}  \psi_R \cr \psi_L  \end{matrix}  \right) =C  \left(  \begin{matrix}  t \cr E - \sqrt{E^2-t^2}  \end{matrix}  \right) e^{i x \sqrt{E^2-t^2}/v } \\[0.5em]
&+ D  \left(  \begin{matrix}  t \cr E + \sqrt{E^2-t^2}  \end{matrix}  \right) e^{-i x \sqrt{E^2-t^2}/v } 
\end{split}
\end{align}
The  last equation follows from looking for the solution in the form of $e^{i q x}$, then finding the eigenvalues of the matrix
\begin{equation} 
\left( \begin{matrix}  v q & t \cr t & - v q \end{matrix} \right) 
\end{equation}\\[0.25em]
which are $E = \pm \sqrt{v^2 q^2 +t^2}$. This gives $qv  = \pm \sqrt{E^2 - t^2}$. Demanding continuity of the wave functions we have
\begin{align}  e^{-i E d/v} &= Ct e^{-i d \sqrt{E^2-t^2} /v} + D t e^{i d \sqrt{E^2-t^2}}, \\[0.5em]
R e^{i E d/v} &= C (E-\sqrt{E^2-t^2}) e^{-i d \sqrt{E^2-t^2}/v} \\ & \hspace{1em} 
+ D (E+\sqrt{E^2-t^2}) e^{i d \sqrt{E^2-t^2}/v}, \nonumber \\[0.5em] 
\vspace{0.7em}
T e^{i E d/v} &= Ct e^{i d \sqrt{E^2-t^2} /v} + D t e^{-i d \sqrt{E^2-t^2}}, \\[0.5em]
\vspace{0.5em}
0 &= C (E-\sqrt{E^2-t^2}) e^{i d \sqrt{E^2-t^2}/v}  \\ & \hspace{1em} + D (E+\sqrt{E^2-t^2}) e^{-i d \sqrt{E^2-t^2}/v}. \nonumber
\end{align}

Solving these gives
\begin{align} & T = \\
&\frac{\sqrt{E^2-t^2} e^{-2 i d E/v}}{\sqrt{E^2-t^2} \cos \left( 2 d \sqrt{E^2-t^2}/v \right) - i E \sin \left( 2 d \sqrt{E^2-t^2}/v \right) } , \nonumber
\end{align}
\begin{align} & R = \\
&- i \frac{t \sin \left( 2 d \sqrt{E^2-t^2}/v \right) \ e^{-2 i d E/v}}{\sqrt{E^2-t^2} \cos \left( 2 d \sqrt{E^2-t^2}/v \right) - i E \sin \left( 2 d \sqrt{E^2-t^2}/v \right) }  \nonumber
\end{align}
\begin{figure}[h!]	\includegraphics[width=7.2 cm]{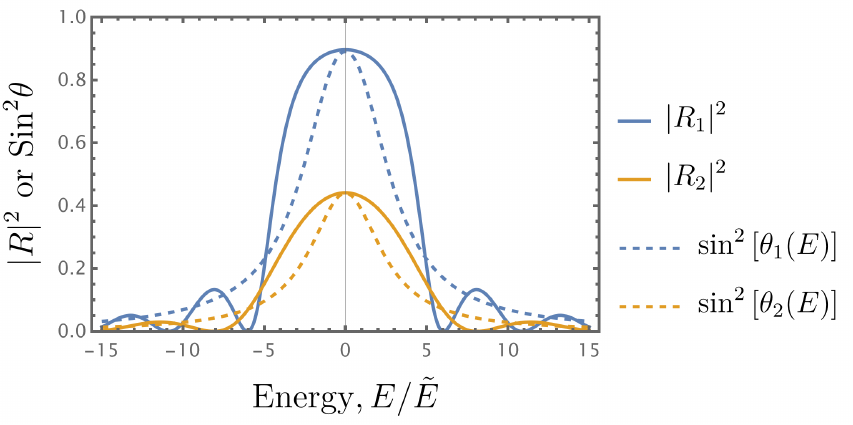}
	\caption{The tunneling probability for the chiral wire calculation $|R^2|$ (solid lines) and for the approximate parameterization used in the band reconstruction $\sin^2 \theta(E)$ (dashed lines) as a function of energy, where $R$ is defined in Eq. \ref{eq:reflectioncoeff} and $\theta(E)$ in Eq. \ref{eq:Ham2}. The energy scale for $|R|^2$ is $\hbar v/d = \hbar (10/3)$ and $|R_1|^2$ ($|R_2|^2$) corresponds to t=2.8 (t=1.5). $\theta_1(E)$ ($\theta_2(E)$) corresponds to $\theta_{max} = \pi/3$ ($\theta_{max} = \pi/6$).}
    \label{fig:tunnelingprob}
\end{figure}

Now if $-t<E<t$, then $qv = \pm i \sqrt{t^2-E^2}$. The equations for $T$ and $R$ remain valid with the substitution $\sqrt{E^2-t^2} \rightarrow i \sqrt{t^2-E^2}$. 
This gives
\begin{align}
&T = \\
&- \frac{\sqrt{t^2-E^2} \   e^{-2 i d E/v}}{\sqrt{t^2-E^2} \cosh \left( 2 d \sqrt{t^2-E^2}/v \right) + i E \sinh \left( 2 d \sqrt{t^2-E^2}/v \right) } 
, \nonumber
\end{align}
\begin{align} \label{eq:reflectioncoeff}
&R =  \\
&\frac{t \sinh \left( 2 d \sqrt{t^2-E^2}/v \right) \ e^{-2 i d E/v} }{i \sqrt{t^2-E^2} \cosh \left( 2 d \sqrt{t^2-E^2}/v \right) + E \sinh \left( 2 d \sqrt{t^2-E^2}/v \right) } . \nonumber
\end{align}
\vspace{0.5em}

$R$ is the amplitude for an incoming $\psi_R$ to go to an outgoing $\psi_L$. This corresponds to the chiral mode which is trapped around one skyrmion tunneling to an adjacent skyrmion. This tunneling probability is plotted as a function of energy in Fig.  \ref{fig:tunnelingprob}.

\end{document}